\documentclass[aps,prb,twocolumn,floatfix,showpacs,superscriptaddress]{revtex4-1}

\usepackage{graphicx,amsmath,amsfonts}

\begin{document}

\title{Intrinsic spin orbit torque in a single domain nanomagnet}

\author{A. Kalitsov}
\affiliation{MINT Center, University of Alabama, Tuscaloosa, AL 35487-0209, USA}
\affiliation{Western Digital Corporation, San Jose, CA 95131, USA}

\author{S. A. Nikolaev}
\affiliation{Univ. Grenoble Alpes, INAC-SPINTEC, F-38000 Grenoble, France} 
\affiliation{CNRS, SPINTEC, F-38000 Grenoble, France} 
\affiliation{CEA, INAC-SPINTEC, F-38000 Grenoble, France}

\author{J. Velev}
\affiliation{Department of Physics, University of Puerto Rico, San Juan, Puerto Rico 00931, USA}
\affiliation{Univ. Grenoble Alpes, INAC-SPINTEC, F-38000 Grenoble, France} 

\author{M. Chshiev}
\affiliation{Univ. Grenoble Alpes, INAC-SPINTEC, F-38000 Grenoble, France} 
\affiliation{CNRS, SPINTEC, F-38000 Grenoble, France} 
\affiliation{CEA, INAC-SPINTEC, F-38000 Grenoble, France}

\author{O. Mryasov$^\dagger$}
\affiliation{MINT Center, University of Alabama, Tuscaloosa, AL 35487-0209, USA}
\affiliation{Western Digital Corporation, San Jose, CA 95131, USA}

\date{\today}

\begin{abstract}
\noindent We present  theoretical studies of the intrinsic spin orbit torque (SOT) in a single domain ferromagnetic layer with Rashba spin-orbit coupling (SOC) using the non-equilibrium Green's function formalism for a tight-binding Hamiltonian. We find that, in the case of a small electric field, the intrinsic SOT to first order in SOC has only the field-like torque symmetry and can be interpreted as the longitudinal spin current induced by the charge current and Rashba field. We analyze the results in terms of the material-related parameters of the electronic structure, such as the band filling, band width, exchange splitting, and the Rashba SOC strength. On the basis of these numerical and analytical results, we discuss the magnitude and sign of SOT. Our results suggest that the different sign of SOT in identical ferromagnets with different supporting layers, e.g. Co/Pt and Co/Ta, can be attributed to electrostatic doping of the ferromagnetic layer by the support.
\end{abstract}

\maketitle

\section{Introduction}

\noindent Magnetization switching in nanoscale devices induced by electric currents has been a subject of intensive research in recent years.\cite{Ohno} One of the most studied approaches is the transfer of spin angular momentum between non-collinear ferromagnetic layers, an effect known as spin transfer torque (STT).\cite{Slonczewski, Berger} In this case, the charge current is spin polarized by a magnetic layer with a pinned magnetization and the spin angular momentum is deposited in a free layer causing precession and reversal of the magnetization. Recently, an alternative way to produce spin torque and manipulate the magnetization direction in a ferromagnetic layer was demonstrated, which it does not require the presence of a second polarizing ferromagnet. Instead, the spin torque is produced by spin-orbit coupling (SOC).\cite{Miron-SOT-1, Miron-SOT-2, Buhrman-SOT-1} This spin-orbit torque (SOT) was observed in $3d$ ferromagnets grown on $5d$ materials with strong SOC, such as Pt\cite{Miron-SOT-1, Miron-SOT-2, Buhrman-SOT-2, Emori-SOT} or Ta.\cite{Buhrman-SOT-1,Miron-SOT-3, Kim-SOT} In this setup, the charge current flows in the plane parallel to the interface and SOT has been shown to produce domain wall motion \cite{Miron-DW, Emori-DW} and magnetization precession.\cite{Demidov-NO,Buhrman-NO} Moreover, recent studies reported on a giant SOT arising at the interface between a topological insulator and a ferromagnet\cite{Ralph-SOT-TI, Hyunsoo} or another magnetically-doped topological insulator.\cite{Tserkovnyak-SOT-TI}

\par The presence of SOT in single nanomagnets has been predicted theoretically based on analytical models.\cite{Manchon-SOT-1,Manchon-SOT-2} Two principal mechanisms of SOT have been proposed. The first mechanism is based on the Rashba effect at the interface between a ferromagnetic layer and supporting non-magnetic metal with strong SOC. The charge current passing through the ferromagnet produces intrinsic torques due to Rashba SOC.\cite{Manchon-SOT-1, Miron-SOT-1, Miron-SOT-2, Pesin, Bijl, Kim} The second mechanism is based on the bulk spin Hall effect (SHE) in the support. In this case, the charge current passing through the support produces spin accumulation at the interface which exerts SOT on the magnetization in a ferromagnet. \cite{Buhrman-SOT-1, Buhrman-SOT-2, Haney-2, Khvalkovskiy} The particular experimental setup determines which of the two mechanisms dominates. The observation of a strong dependence of the SOT magnitude on the support thickness and its sign reversal at small thicknesses\cite{Kim-SOT} suggest that for a thick supporting layer the current flows predominantly through the support and the SHE-SOT dominates. On the other hand, when the support is very thin the current flows through the ferromagnetic layer and the Rashba-SOT dominates. 

\par Many theoretical works employ a picture of conduction electrons interacting with localized magnetic moments in the presence of Rashba SOC\cite{Manchon-SOT-1, Pesin, Bijl, Kim} or SHE.\cite{Haney-2, Khvalkovskiy} The conduction electrons are assumed to be free and their interaction with localized moments is treated on the level of the $s$-$d$ exchange model.\cite{Schubin,Zener} In the context of transport calculations, the majority of works deal with the quasiclassical Boltzmann approach.\cite{Manchon-SOT-1, Bijl, Haney-2, Khvalkovskiy,Manchon-SOT-3}  This assumes a linear regime, in which SOT is proportional to the charge current and the torquance is expressed in terms of phenomenological parameters, such as the spin Hall angle and spin polarization of the carriers.  There is also a couple of reports on first-principles calculations of SOT, where the torquance is related to the Berry phase curvature of the occupied states.\cite{Haney-3, Freimuth} Nevertheless, the band structure dependence and the finite bias behavior of SOT remain largely unexplored. 

\par In this paper, we discuss the intrinsic SOT arising from the band structure alone for a ferromagnetic layer with Rashba spin-orbit coupling. We develop ballistic transport formalism based on the Keldysh non-equilibrium Green's function (NEGF) method and a single-orbital tight-binding (TB) Hamiltonian model. To gain insights into physical picture, we derive an analytic expression for SOT to first order in SOC and use it to analyze the SOT dependence on the band structure parameters and applied voltage. We show that keeping only the first order terms in SOC leads to field-like SOT in the ballistic regime.

\section{Methodology}

\noindent We consider a two dimensional ferromagnetic layer in the so-called current-in-plane (CIP) geometry schematically shown in Fig.~1$a$. The direction of the charge current $\boldsymbol{j}$ is chosen to be along the $x$ axis and the unit vector $\boldsymbol{S}$ along the magnetization $\boldsymbol{M}$ is given in the conventional spherical coordinate system, $\boldsymbol{S}=(\cos{\phi}\sin{\theta},\sin{\phi}\sin{\theta},\cos{\theta})$. The induced SOT can be separated into two components which have the symmetry of antidamping-like torque (DLT), $\boldsymbol{T}_{\parallel} = T_{\parallel}\,\boldsymbol{S}\times[(\boldsymbol{e}_{z}\times\boldsymbol{j})\times\boldsymbol{S}]$, and field-like torque (FLT), $\boldsymbol{T}_{\perp} = T_{\perp} (\boldsymbol{e}_{z}\times\boldsymbol{j})\times\boldsymbol{S}$. These torques are also referred to as parallel and perpendicular to the plane given by the directions of the magnetization $\boldsymbol{S}$ and  $(\boldsymbol{e}_{z}\times\boldsymbol{j})$, where $\boldsymbol{e}_{z}$ is the unit vector along the $z$ axis.

\begin{figure}
\begin{center}
\includegraphics[width=0.48\textwidth]{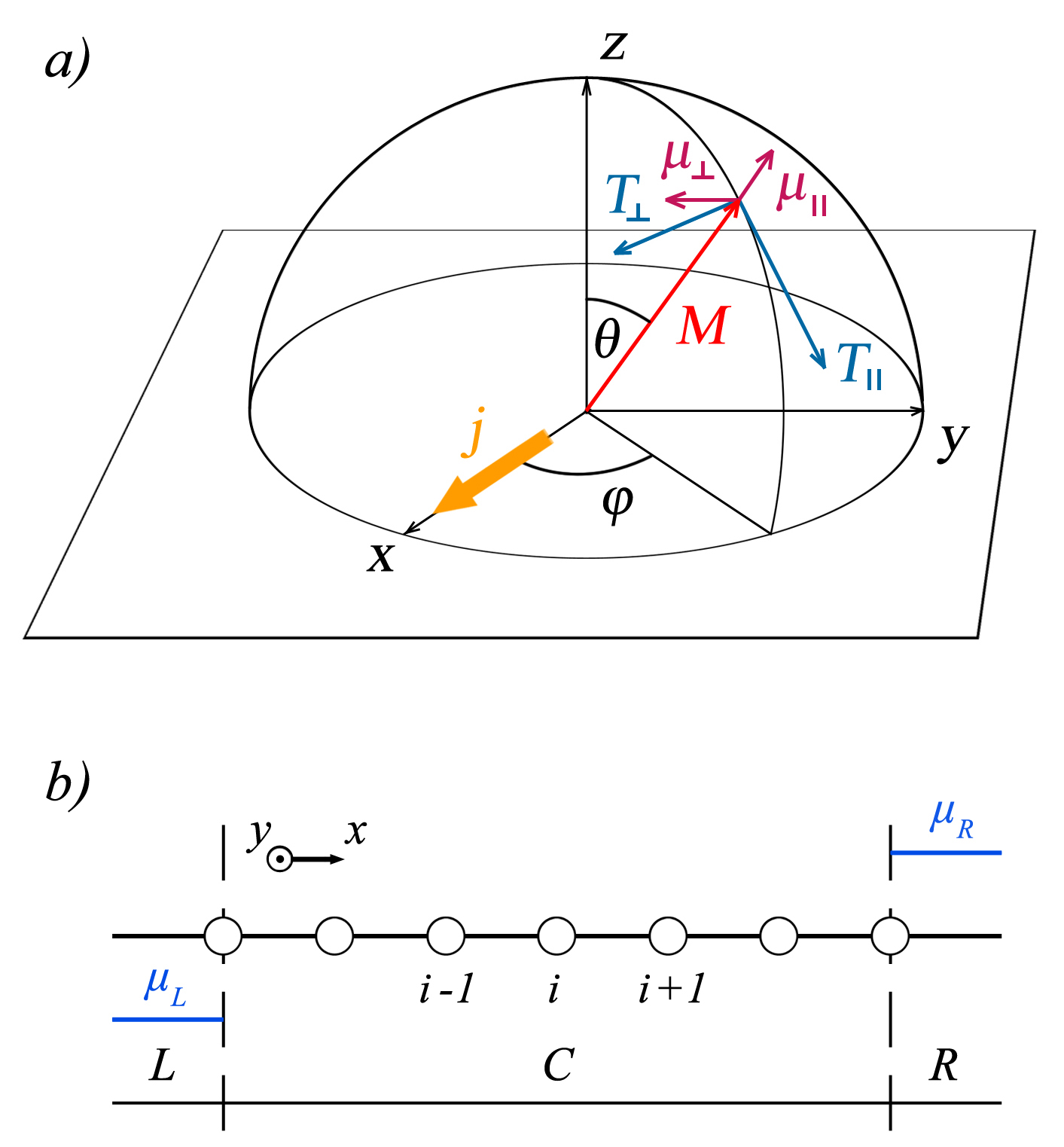}
\end{center}
\caption{$a)$ Schematic view of the 2D ferromagnetic layer in the $xy$ plane. The charge current $\boldsymbol{j}$ direction is along the $x$-axis. The magnetization $\boldsymbol{M}$ is defined by the spherical angles $\theta$ and $\phi$. The purple arrows indicate the current-induced spin density. The blue arrows denote $\boldsymbol{T}_{\perp}$ and  $\boldsymbol{T}_{\parallel}$ contributions to the total SOT. $b)$ Schematic view of the two-probe setup consisting of the scattering region $C$ and two electrodes $L$ and $R$. }
\end{figure}

\subsection{Hamiltonian matrix elements}

\noindent The Hamiltonian of the system in the absence of SOC is 
\begin{equation}
\hat{\mathcal{H}}_{0} = \sum_{nm,\sigma} t_{nm} \hat{c}_{n}^{+\sigma} \hat{c}_{m}^{\sigma} - j_{ex} \sum_{n,\sigma\sigma'} \hat{c}_{n}^{+\sigma'} (\boldsymbol{\sigma} \cdot \boldsymbol{S})^{\sigma'\sigma} \hat{c}_{n}^{\sigma} \label{Eq-Ho},
\end{equation}
\noindent where the first term corresponds to the conduction electrons with diagonal elements representing the onsite energies, $\varepsilon_{0} = t_{nn}$, and off-diagonal elements representing the electron hopping parameters, $t_{n,m \ne n}$. We consider the hopping parameter to be non-zero between nearest neighbors only, $t = t_{n,n+1}=t_{n+1,n}$. The second term represents the $s$-$d$ model, where $j_{ex}$ stands for the exchange coupling and $\boldsymbol{\sigma}$ is the vector of the Pauli matrices.\cite{Schubin,Zener} The Rashba SOC Hamiltonian is written in the tight-binding basis as \cite{Pareek}
\begin{equation}
\begin{aligned}
\hat{\mathcal{H}}_{SO} &= -\lambda \sum_{n,\sigma\sigma'} \left[ \hat{c}^{+\sigma'}_{n +\boldsymbol{e}_{y}} (i\sigma_{x}^{\sigma'\sigma}) \hat{c}^{}_{n}\right.\\
&\left. \qquad\qquad -\hat{c}^{+\sigma'}_{n+\boldsymbol{e}_{x}} (i\sigma_{y}^{\sigma'\sigma}) \hat{c}_{n}^{} +  \mathrm{H.C.} \right],
\end{aligned} \label{Eq-Hsoc}
\end{equation}
\noindent where $\lambda$ is the Rashba SOC strength, $\boldsymbol{e}_{x}$  and $\boldsymbol{e}_{y}$ are the unit vectors in the $x$ and $y$ axes respectively. 

\par The equations above can be Fourier transformed and written in the momentum space as
\small
\begin{equation}
\hat{\mathcal{H}}_{0}(\boldsymbol{k}) = \left( \begin{array}{cc}
\varepsilon_{0} - j_{ex} \cos{\theta} + t \, p(\boldsymbol{k}) & - j_{ex} \sin{\theta} \,  e^{-i\phi} \\
-j_{ex} \sin{\theta} \,  e^{i\phi} & \varepsilon_{0} + j_{ex} \cos{\theta} + t \, p(\boldsymbol{k})  \end{array} \right)
\end{equation}
\normalsize
and
\begin{equation}
\hat{\mathcal{H}}_{SO}(\boldsymbol{k}) = -\lambda \left( \begin{array}{cc}
0 & r(\boldsymbol{k}) \\
r^{*}(\boldsymbol{k}) & 0  \end{array} \right) \label{eq-H},
\end{equation} 
where $p(\boldsymbol{k})=2(\cos{k_{x}}+\cos{k_{y}})$ and $r(\boldsymbol{k})=2(i\sin{k_{x}}+\sin{k_{y}})$. Finally, the total Hamiltonian of the system is $\hat{\mathcal{H}} = \hat{\mathcal{H}}_{0} + \hat{\mathcal{H}}_{SO}$.

\subsection{Non-equilibrium charge and spin transport for current-in-the-plane geometry}

\noindent In order to calculate transport properties we separate the system into three regions along the $x$ axis, a scattering region (C) connected to the left (L) and right (R) semi-infinite leads all made of the same material as shown in Fig.~1$b$.  A finite voltage drop $e \mathrm{V} = \mu_{L}  -\mu_{R}$ is introduced between the leads by maintaining them in local thermodynamic equilibrium with the corresponding chemical potentials $\mu_{L(R)}$. Thus, the occupation of the leads is governed by the Fermi-Dirac distribution function $f_{L(R)} = 1 / (1 + e^{(E-\mu_{L(R)})/k_{B}T})$. The system is assumed to be periodic in the $y$ direction perpendicular to the current. For a fixed applied voltage, the choice of size of the scattering region determines the electric field in the material. In the limit of an infinite scattering region the electric field tends to zero.

\par In this setup, the charge and spin current densities between any two neighboring planes $i$ and $i+1$ in the $x$ direction are given in the NEGF formalism as \cite{Kalitsov-STT}
\begin{equation}
\begin{aligned}
I = \frac {e} {\hbar} \int \frac {dE\, dk_{y}} {4 \pi^{2}} \, & \mathrm{Tr}_{\sigma} [ \hat{\mathcal{H}}_{i,i+1} \hat{G}^{<}_{i+1,i}(E,k_{y}) \\
& - \hat{\mathcal{H}}_{i+1,i} \hat{G}^{<}_{i,i+1}(E,k_{y}) ]
\end{aligned} \label{eq-I}
\end{equation} 
\noindent and
\begin{equation}
\begin{aligned}
\boldsymbol{I}^{S} = \int \frac {dE\, dk_{y}} {8 \pi^{2}} \, & \mathrm{Tr}_{\sigma} [ \hat{\mathcal{H}}_{i,i+1} \hat{G}^{<}_{i+1,i}(E,k_{y}) \\
& - \hat{\mathcal{H}}_{i+1,i} \hat{G}^{<}_{i,i+1}(E,k_{y})) \boldsymbol{\sigma} ],
\end{aligned} \label{eq-Is}
\end{equation}  
\noindent where all quantities are $2 \times 2$ matrices in the spin space, $\mathcal{H}_{i,i+1}$ is the Hamiltonian matrix element between the planes $i$ and $i+1$, and $G^{<}_{i,i+1}$ is the NEGF. The integration is performed over the energy and $y$ component of the wave vector $\boldsymbol{k}$. 

\par Similarly the spin density of conduction electrons arising from the $s$-$d$ exchange coupling is given by \cite{Kalitsov-STT-1}
\begin{equation}
\boldsymbol{\mu} = - i \mu_{B} \int \frac {dE\, dk_{y}} {8 \pi^{2}} \, \mathrm{Tr}_{\sigma} \left[ \hat{G}_{i,i}^{<}(E,k_{y}) \boldsymbol{\sigma} \right],  \label{eq-M}
\end{equation} 
\noindent where $\mu_{B}$ is the Bohr magneton. All quantities are independent of the plane index $i$. Having calculated the magnetic moment, the spin torque can be readily found as
\begin{equation}
\boldsymbol{T} = -\frac{j_{ex}} {\mu_{B}} \,\boldsymbol{S} \times \boldsymbol{\mu}. \label{eq-T}
\end{equation}

\subsection{Non-equilibrium Green's function matrix elements}
\noindent The NEGF for a standard two-probe geometry is written in the form \cite{Datta, Lifshitz}
\begin{equation}
\hat{G}^{<} = i \hat{G} (f_{L} \hat{\Gamma}_{L} + f_{R} \hat{\Gamma}_{R}) \hat{G}^{+} \label{eq-NEGF},
\end{equation}  
\noindent where $\hat{G}$ is the retarded Green's function (GF), $\hat{\Gamma}_{L(R)} = i (\hat{\Sigma}_{L(R)} - \hat{\Sigma}^{+}_{L(R)})$ is the escape rate to the electrodes, $\hat{\Sigma}_{L(R)} = \hat{\mathcal{H}}_{L(R)C}^{+} \hat{g}^{}_{LL(RR)} \hat{\mathcal{H}}^{}_{L(R)C}$ is the self-energy due to the attachment of electrodes, and $\hat{g}_{LL(RR)}^{}$ is the surface GF of each electrode. Using this expression, the non-equilibrium spin density Eq.~(\ref{eq-M}) and SOT Eq.~(\ref{eq-T}) can be calculated in general, for any arbitrary strength of Rashba SOC, size of the scattering region and applied electric field. The general expression for SOT given by Eq.~(\ref{eq-M}), however, obscures the fundamental physics of the phenomenon.
\par In order to elucidate the behavior of SOT and derive its analytical expression, we simplify our further approach by adopting the limiting case of large scattering region where the bias induced electric field almost vanishes. Moreover, due to the periodicity along the $y$ axis, the retarded GF matrix elements can be written in the mixed (real and reciprocal space) representation as the Fourier transform of GF, $\hat{G}(\boldsymbol{k})$, in  momentum space
\begin{equation}
\hat{G}_{n,m}(k_{y}) = \frac{1} {2 \pi} \, \int \limits_{-\pi}^{\pi} dk_{x} \, e^{ik_{x}(n-m)} \hat{G}_{nm} (\boldsymbol{k}),
\end{equation}
\noindent where $\hat{G}(\boldsymbol{k}) = \left[  (E + i \eta) \hat{I} - \hat{\mathcal{H}}(\boldsymbol{k}) \right]^{-1}$, $\eta$ is a positive infinitesimal, $n$ and $m$ are layer indexes along the $x$ axis. Under these assumptions, the surface GF of the electrodes can be found from the bulk GF, because connecting two semi-infinite systems restores an infinite periodic system.\cite{Velev} For the self-energies we obtain
$\hat{\Sigma}_{L} = \hat{G}^{-1}_{i+j,i+1} \hat{G}_{i+j,i}^{} \hat{\mathcal{H}}_{i,i+1}^{}$ and 
$\hat{\Sigma}_{R} = \hat{G}^{-1}_{i,i+j} \hat{G}_{i,i+j+1}^{} \hat{\mathcal{H}}_{i+1,i}^{}$,
where $i\in(-\infty,\infty)$ and $j\in[1,\infty)$. Since the bulk GF is periodic, its matrix elements depend only on the difference of their coordinates, $\hat{G}_{n,m} = \hat{G}_{n-m}$. Thus, substituting into Eq.~(\ref{eq-NEGF}) we obtain
\small
\begin{equation}
\begin{aligned}
\hat{G}_{i,j}^{<} = & - f_{L} \hat{G}_{i+1,i}^{} \hat{\mathcal{H}}_{i,i+1}^{} \hat{G}^{+}_{i,i} + f_{L} \hat{G}_{i,i}^{} \hat{\mathcal{H}}_{i+1,i}^{} \hat{G}^{+}_{i,i+1} \\
& - f_{R} \hat{G}_{i,i+1}^{} \hat{\mathcal{H}}_{i+1,i}^{} \hat{G}^{+}_{i,i} + f_{R} \hat{G}_{i,i}^{} \hat{\mathcal{H}}_{i,i+1}^{} \hat{G}^{+}_{i+1,i}.
\end{aligned}
\label{eq:ginf}
\end{equation}
\normalsize 
\noindent Finally, taking into account an explicit form of the Hamiltonian matrix elements $\hat{\mathcal{H}}_{i,i+1} = \hat{I} t - i \lambda \hat{\sigma}_{y}$ and using $\hat{\mathcal{H}}^{+}_{i+1,i}=\hat{\mathcal{H}}_{i,i+1}^{}$, the diagonal NEGF matrix element can be written as
\begin{equation}
\begin{aligned}
\hat{G}^{<}_{i,i} & = f_{L}  t  ( \hat{G}_{i,i}^{} \hat{G}_{i,i+1}^{+} - \hat{G}_{i+1,i} ^{}\hat{G}^{+}_{i,i} ) \\
& + f_{R} t ( \hat{G}_{i,i}^{} \hat{G}^{+}_{i+1,i} - \hat{G}_{i,i+1}^{} \hat{G}^{+}_{i,i} ) \\
& + i f_{L} \lambda ( \hat{G}_{i,i}^{} \hat{\sigma_{y}} \hat{G}_{i,i+1}^{+} + \hat{G}_{i+1,i}^{} \hat{\sigma}_{y} \hat{G}_{i,i}^{+} ) \\
& - i f_{R} \lambda ( \hat{G}_{i,i+1}^{} \hat{\sigma_{y}} \hat{G}_{i,i}^{+} + \hat{G}_{i,i}^{} \hat{\sigma}_{y} \hat{G}_{i+1,i}^{+}). \\
\end{aligned} \label{eq-NEGF-final}
\end{equation}  

\par   Now we can expand the NEGF matrix elements in orders of the SOC strength and retain in the expressions only the lowest order in $\lambda$. The somewhat lengthy algebra is given in Appendix A. The principal result is that the spin density can be decomposed into two components, as follows
\begin{equation}
\boldsymbol{\mu} = \boldsymbol{\mu}_{\parallel} + \boldsymbol{\mu}_{\perp} = \boldsymbol{S} (\mu_{0} +2 S_{y} \mu_1) + (0, -2 \mu_1, 0),
\end{equation}
where $\mu_{0}$ and $\mu_1$ are the zeroth and first orders in the expansion in $\lambda$. From Eq. (\ref{eq-T}) it is clear that only the second term produces SOT, $\boldsymbol{T} = T_{\perp} \, (S_{z}, 0, -S_{x})$, which has the field-like symmetry, where
\begin{equation}
T_{\perp} = -\frac{2j_{ex}\mu_1}{\mu_{B}} = \frac{\hbar\lambda}{et} (I^{\uparrow} - I^{\downarrow}) = \frac{2\lambda}{t} I^{S_{z}}.
\label{eq-T-approx}
\end{equation}
\noindent  This field-like SOT can be interpreted as the longitudinal spin current induced by the charge current and SOC. Taking into account that the spin current is an even function with respect to the time-reversal symmetry, this expression is an odd function of the magnetization. Moreover, for an infinite scattering region the damping-like torque is expected to appear to second order in the SOC strength, i.e. $T_{\perp} \gg T_{\parallel}$. Here, it is worth to note that for a finite size system, when an applied electric field has to be taken into account, the corresponding solution of Eq.~(\ref{eq-NEGF}) reveals an additional DLT component. However, as the size of the scattering region increases, this DLT vanishes, while the FLT survives and converges to a form given by Eq.~(\ref{eq-T-approx}). Thus, we state that for a finite size system where the bias induced voltage drop can not be neglected the resulting SOT exhibits both the FLT and DLT components, which can be regarded of intrinsic origin in the ballistic limit.
\par Similar expression for the FLT was obtained in the context of the Boltzmann transport equation.\cite{Manchon-SOT-1} The main difference comes from the underlying model and method. In the Boltzmann approach longitudinal spin current is proportional to the momentum relaxation time, which diverges at low impurity concentrations and gives unphysical infinite current and torque, while our approach based on the Keldysh formalism shows the correct ballistic limit. Our results are in agreement with recent theoretical predictions based on the non-equilibrium Green's function formalism showing that low charge currents flowing solely at the interface of a ferromagnetic layer and topological insulator can induce antidamping-like SOTs.\cite{branislav} In addition, the Keldysh formalism is a completely general approach and can be extended to include any kind of disorder.\cite{Kalitsov-CPA} In this case, the impurity contribution to SOT would be extrinsic and depend on the impurity type and their distribution.
\par Our result is also in agreement with previous studies which show that the interfacial SOC results predominantly in field-like torques.\cite{Haney-3,Freimuth} However, some recent publications reported on the intrinsic DLT of a comparable magnitude.\cite{Manchon-SOT-3,antidamp1} In this approach based on the linear response theory, SOT is analyzed in terms of the intraband and interband electronic transitions arising due to the effect of the electric field on the charge distribution and the shape of wave functions, respectively. These studies revealed that the only intrinsic component is the DLT, while the FLT is of extrinsic origin as it is inversely proportional to the spectral broadening caused by impurities. Interestingly, other complementary studies showed that the DLT cancels when the vertex corrections are taken into account.\cite{Titov} We believe that the linear response theory has its well known drawbacks, as it also suffers in the clean limit leading to a divergence of FLT. In this regard, the Keldysh formalism is a more reliable tool that allows us to investigate the origin of SOT even in the ballistic regime. As it is clear from our derivation, for example, Eqs.~(\ref{eq:additional}) and (\ref{eq-nefg-matrix}), the interband transitions disappear once the voltage drop is neglected, so only the intraband transitions survive giving rise to the FLT. In case of a finite electric field present in the scattering region, Eq.~(\ref{eq-NEGF}) contains additional spin mixing terms, which correspond to the intrerband transitions in our single-orbital model and result in appearance of the DLT component.

\section{RESULTS AND DISCUSSION}

\begin{figure}
\begin{center}
\includegraphics[width=0.4\textwidth]{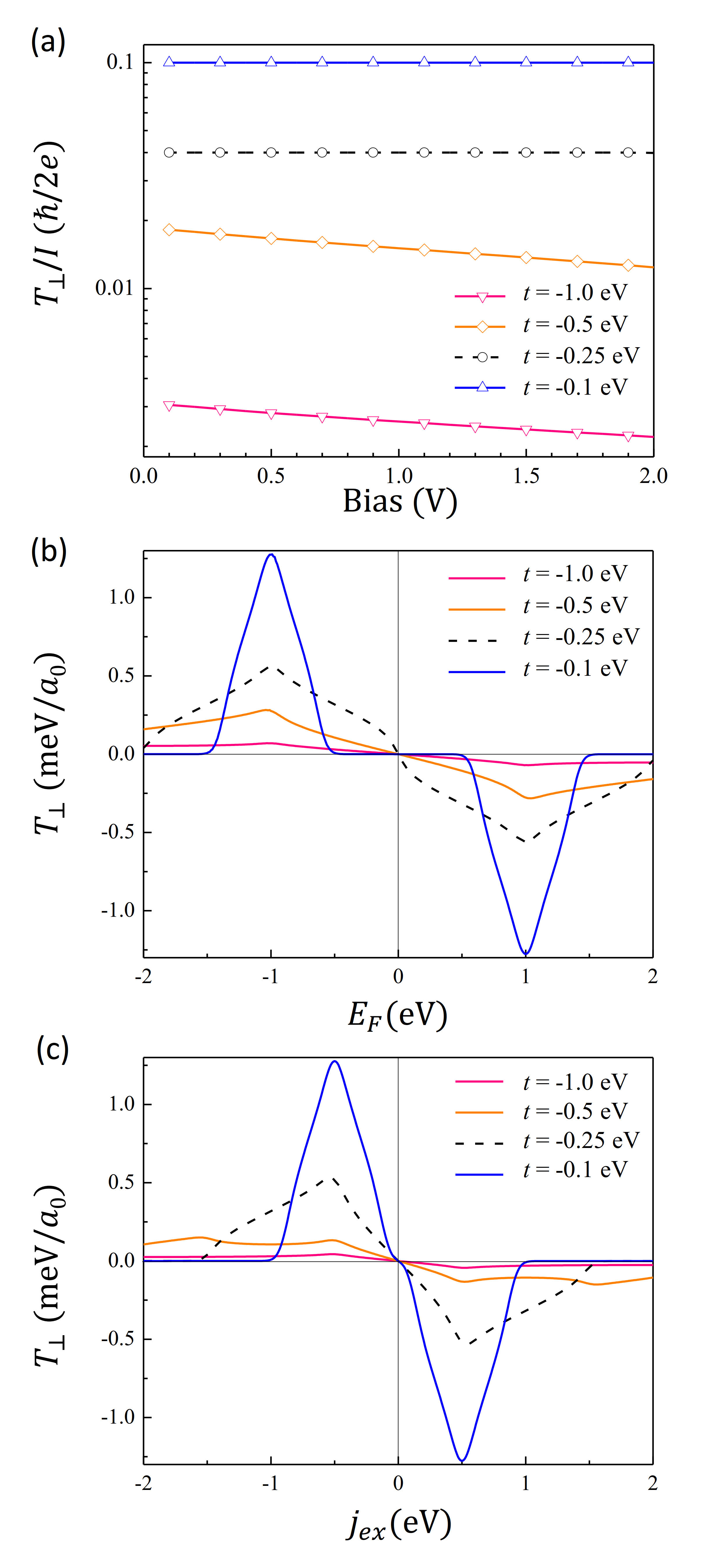}
\end{center}
\caption{$a)$ Bias dependence of the SOT efficiency  for different values of the electron hopping parameter $t$ ($j_{ex} = 1.0$ eV, $\lambda = 0.01$ eV and $E_{F} = 1.0$ eV); $b)$ SOT dependence on the Fermi level ($j_{ex}  =1.0$ eV, $\lambda = 0.01$ eV and $\mathrm{V} = 0.1$ V); $c)$ SOT dependence on the exchange interaction parameter for different values of the electron hopping parameter ($E_{F} = 1.0$ eV, $\lambda = 0.01$ eV and $\mathrm{V} = 0.1$ V). Here, $a_{0}$ indicates a unit cell length and the negative sign of $j_{ex}$ stands for the magnetization reversal.}
\end{figure}

\noindent The first step to perform calculations of SOT is a parametrization of the tight-binding model. Many experimental studies use Co or CoFe as the ferromagnetic layer and Pt or Ta as the support.\cite{Manchon-SOT-1,Buhrman-SOT-1,Miron-SOT-3} Thus, the ferromagnetic material is either Fe or Co, which have very similar band structures but differ in occupation by one electron. Within the $s$-$d$ exchange model, centers of the majority and minority bands are at $\varepsilon^{\uparrow} = \varepsilon_{0} - j_{ex}$ and $\varepsilon^{\downarrow} = \varepsilon_{0} + j_{ex}$, respectively. The exchange splitting between two bands is $2j_{ex}$ and the band width is $8t$. Thus, for $|j_{ex}/t| > 4$ the material has a gap between the majority and minority bands. For a typical ferromagnet, such as Co or Fe, the exchange splitting is of the order of a couple of eV and the band width is of the order of several eV. Therefore, this band structure can be described by choosing $j_{ex} = 1.0$\,eV and $t = -1.0$\,eV, respectively, which is representative for a metallic ferromagnet with the exchange splitting of 2.0\,eV. Note that for $j_{ex} = 1.0$\,eV, the case of $|t| < 0.25$\,eV is an insulator. Without any loss of generality, we set the onsite energy $\varepsilon_{0} = 0.0$\,eV,  while the band occupation is controlled by shifting the Fermi level $E_F$.  We also choose $\lambda = 0.01$ eV for the SOC strength.

\par Using this parametrization, we calculate SOT from Eq.~(\ref{eq-T-approx}) and plot in Fig.~2$a$ the SOT efficiency as a ratio of SOT to the charge current, $T_{\perp}/I$. The Fermi level is chosen at $E_{F} = -1.0$\,eV, which corresponds to the half-filled majority band and completely empty minority band. For comparison between the metallic ($t = -1.0$\,eV) and insulating ($t = -0.1$\,eV) cases, we also plot the SOT efficiency for several different band widths. The main observation is that the SOT efficiency decreases exponentially with the band width. Another observation is that SOT is fairly independent of the bias. Because of the choice of $E_F$, only the majority band contributes to the transport, $I^{\uparrow}\ne 0$ and $I^{\downarrow}=0$, as a result SOT is proportional to the current and the SOT efficiency is constant. At higher bias, the minority band also contributes to the charge current, and the overall efficiency and SOT decrease when the current increases.

\par In order to gain further insight into the origin of SOT we investigate its dependency on the main parameters of the model, i.e. band filling and exchange splitting. In Fig.~2$b$ the dependence on the band occupation is shown for different band widths. The applied bias voltage is set to $\mathrm{V} = 0.1$ $V$. As it is seen, SOT is an antisymmetric function with respect to $E_{F}$. The magnitude of $T_{\perp}$ peaks around the middle of the majority and minority bands at $\varepsilon_{0} - E_{F} = \mp 1.0$\,eV, respectively. The contributions to SOT come only from the energy regions with available carriers, therefore, for an insulating system SOT has two narrow peaks around the band centers. We find that partially occupied bands contribute to  SOT, with less than half-filled majority bands producing positive contribution and more than half-filled negative and vice versa for the minority bands. Therefore, the maximum SOT values correspond to the $1/4$ and $3/4$ filled bands. 

\par It is worth noting that in the area of $1/2$ filling the sign of $T_{\perp}$ can change for small charge doping shifting of $E_F$ to the left or right. This fact can explain the difference in the sign of SOT for the cases of Co/Pt and Co/Ta.\cite{Miron-SOT-3} Chemically, Co (or CoFe) is between Ta and Pt in electronegativity. Interfacing the ferromagnetic layer with a substrate is essentially chemical doping of the interface. The difference in electronegativity between the ferromagnet and the support will cause charge to flow through the interface, where the flow will be in different direction in the case of the two substrates, holes in the case of Pt and electrons in the case of Ta. Since the ferromagnetic layer is extremely thin this interface doping could change significantly the Fermi level position of the whole system. This could be sufficient to make the transport majority- or minority-dominated, that, in turn, changes the sign of SOT.

\par Calculations with realistic band structures show that the magnitude and sign of the Rashba SOC strength, and respectively the band contribution to SOT, vary from band to band.\cite{Park} Nevertheless, we believe that our model captures the basic physics of the phenomenon. The main observation is that bands with different spin chirality and different occupation give contributions of different signs to SOT. Our simplified band structure model has a particle-hole symmetry which is the reason that SOT changes its sign exactly at 1/2 filling. The sign of SOT in the realistic band structure case will depend on the detailed balance of the occupation of the majority and minority bands. Nevertheless, it is feasible that this balance is changed significantly by means of the interface charge transfer to alter the sign of SOT. Moreover, the Rashba parameters themselves depend on the electric field at the interface, which is determined by the band alignment at the interface. 

\par Finally, in Fig.~2$c$ the dependence of $T_{\perp}$ on the exchange splitting is shown. As it is seen, SOT is a non-monotonous function of $j_{ex}$, that is not obvious, as intuitively it is expected that larger exchange couplings lead to larger SOT. However, SOT reaches its maximum (minimum) value at the same $j_{ex}$ for all curves. This value of $j_{ex}$ corresponds to the case when one of the spin-channels gives the highest contribution.

\section{CONCLUSIONS}

\noindent We present analytical and numerical results for SOT in a single domain ferromagnetic layer with Rashba SOC. We find that, in the limit of large samples, to first order in SOC this torque has the FLT symmetry and is proportional to the longitudinal component of the spin-current, while the DLT component arises for finite size systems due to the bias induced voltage drop. We give the analytical expression of SOT in terms of the material related parameters of the electronic structure that enables physically transparent analysis. Our results indicate that the SOT efficiency decreases with the band width and the magnitude and sign of the band contributions to SOT depend on the band spin component and occupation. This makes it possible to change the overall sign of SOT by electrostatic doping. Thus, experimental observations of the opposite signs of SOT in Co/Pt and Co/Ta layers could be explained in terms of the hole and electron doping of the Co layer from the supporting Pt or Ta layers, respectively. We expect that our results may be useful to selecting specific material combinations with optimal properties.

\begin{acknowledgements}
\noindent The work at UA was supported by C-SPIN, one of the six centers of STARnet, a Semiconductor Research Corporation program, sponsored by MARCO and DARPA. The work at UPR was supported by the National Science Foundation (Grants Nos. EPS-1002410 and EPS-1010094). The work at Spintec was supported by ``Emergence et partenariat strategique" program of Univ. Grenoble Alpes. The authors thank I.~M.~ Miron, G.~Gaudin, S.~Emori, A. Manchon, J. Fabian, W.~H.~Butler and B.~K.~Nikolic for their fruitful comments.

\end{acknowledgements}

\appendix
\section{ Perturbation expansion in SOC parameter}

\noindent If we consider the zeroth order with respect to the SOC parameter, Eq. (\ref{eq-NEGF-final}) can be simplified as
\begin{equation}
\begin{aligned}
\hat{G}^{<(0)}_{i,i} & = f_{L} t ( \hat{g}_{i,i}^{} \hat{g}_{i,i+1}^{+} - \hat{g}^{}_{i+1,i} \hat{g}^{+}_{i,i} ) \\
& + f_{R} t ( \hat{g}_{i,i}^{} \hat{g}^{+}_{i+1,i} - \hat{g}^{}_{i,i+1} \hat{g}^{+}_{i,i}),
\end{aligned}
\end{equation}   
\noindent where $\hat{g}$ is the GF in the absence of SOC. Its explicit angular dependence is given by
\begin{equation}
\begin{aligned}
g^{\uparrow\uparrow} & = \frac{1}{2} ( g^{\uparrow} ( 1 + \cos{\theta}) + g^{\downarrow} (1 - \cos{\theta}) ), \\
g^{\downarrow\downarrow} & = \frac{1}{2} ( g^{\uparrow} ( 1 - \cos{\theta} ) + g^{\downarrow} ( 1 + \cos{\theta}) ), \\
g^{\uparrow\downarrow} & = \frac{1}{2} ( g^{\uparrow} - g^{\downarrow}) \sin{\theta} e^{-i\phi}, \\
g^{\downarrow\uparrow} & = \frac{1}{2} (g^{\uparrow} - g^{\downarrow}) \sin{\theta} e^{i\phi}, \\
\end{aligned} \label{Eq-g-col}
\end{equation}
\noindent and $g^{\uparrow(\downarrow)}$ is the GF corresponding to the case when the magnetization is perpendicular to the plane (or in the local coordinate frame aligned with the magnetization~$\boldsymbol{S}$). For $t < 0$ its analytical expression is
\begin{equation}
\begin{aligned}
g^{\sigma}_{i,j}(k_{y})&=\frac{1}{2\pi}\int\limits_{-\pi}^{\pi}dk_{x}\,\frac{\,e^{ik_{x}(x_{i}-x_{j})}}{E^{\sigma}-2t\cos{k_{x}}+i\eta}\\
&=-i\frac{\left(\frac{E^{\sigma}}{2t}+i\sqrt{1-\left(\frac{E^{\sigma}}{2t}\right)^{2}}\right)^{|x_{i}-x_{j}|}}{\sqrt{4t^{2}-E^{\sigma2}}},
\end{aligned}
\label{eq:gbulk}
\end{equation}
\noindent where $E^{\uparrow(\downarrow)}  =E-\varepsilon_{o} \pm j_{ex} - 2 t \cos{k_{y}}$. The matrix elements of $\hat{G}^{<(0)}_{i,i}$ are then given by
\begin{equation}
\begin{aligned}
G_{i,i}^{<\uparrow\uparrow(0)}=&-it(f_{L}+f_{R})\mathrm{Im}[(g^{\uparrow}_{i+1,i}g^{\uparrow*}_{i,i}+g^{\downarrow}_{i+1,i}g^{\downarrow*}_{i,i})\\
&+(g^{\uparrow}_{i+1,i}g^{\uparrow*}_{i,i}-g^{\downarrow}_{i+1,i}g^{\downarrow*}_{i,i})\cos{\theta}],\\
G_{i,i}^{<\downarrow\downarrow(0)}=&-it(f_{L}+f_{R})\mathrm{Im}[(g^{\uparrow}_{i+1,i}g^{\uparrow*}_{i,i}+g^{\downarrow}_{i+1,i}g^{\downarrow*}_{i,i})\\
&-(g^{\uparrow}_{i+1,i}g^{\uparrow*}_{i,i}-g^{\downarrow}_{i+1,i}g^{\downarrow*}_{i,i})\cos{\theta}],\\
G_{i,i}^{<\uparrow\downarrow(0)}=&-it(f_{L}+f_{R})\mathrm{Im}[(g^{\uparrow}_{i+1,i}g^{\uparrow*}_{i,i}\\
&-g^{\downarrow}_{i+1,i}g^{\downarrow*}_{i,i})\sin{\theta}e^{-i\phi}],\\
G_{i,i}^{<\downarrow\uparrow(0)}=&-it(f_{L}+f_{R})\mathrm{Im}[(g^{\uparrow}_{i+1,i}g^{\uparrow*}_{i,i}\\
&-g^{\downarrow}_{i+1,i}g^{\downarrow*}_{i,i})\sin{\theta}e^{i\phi}].
\end{aligned} \label{Eq-G-col}
\end{equation}
\noindent In the ballistic regime, when the scattering region is large enough, we can safely neglect the voltage drop induced by an applied bias voltage. As it was pointed out, in this limit Eq.~(\ref{eq-NEGF}) is transformed to Eq.~(\ref{eq:ginf}). However, it is worth noting that when dealing with finite size systems the voltage drop should appear in the definition of GFs of Eq.~(\ref{eq-NEGF}) rendering this transformation impossible. Moreover, under this assumption one can further simplify:
\begin{equation}
\mathrm{Im} (g^{\uparrow}_{i+1,i} g^{\uparrow*}_{i,i} - g^{\downarrow}_{i+1,i} g^{\downarrow*}_{i,i}) =-\frac{1}{2t}\mathrm{Im} (g^{\uparrow}_{i,i} -g^{\downarrow}_{i,i} ).
\label{eq:additional}
\end{equation}
\par The magnetic moment arising due to the $s$-$d$ exchange coupling in the absence of SOC can be written as
\begin{equation}
\begin{aligned}
\boldsymbol{\mu}_{0} & = - \mu_{B} t \int \frac{dE\,dk_{y}}{4\pi^{2}}(f_{L}+f_{R})\\
& \,\,\,\,\,\, \times \mathrm{Im} (g^{\uparrow}_{i+1,i} g^{\uparrow*}_{i,i} - g^{\downarrow}_{i+1,i} g^{\downarrow*}_{i,i}) \boldsymbol{S} \\
& = \mu_{B} \int \frac{dE\,dk_{y}}{8\pi^{2}} (f_{L} + f_{R}) \, \mathrm{Im} (g^{\uparrow}_{i,i} - g^{\downarrow}_{i,i}) \boldsymbol{S}.
\end{aligned}
\end{equation}
\noindent This magnetic moment is collinear to the magnetization and, therefore, it does not create any torques. In equilibrium $f_{L} = f_{R} = f(E_{F})$, and we obtain
\begin{equation}
\begin{aligned}
\boldsymbol{\mu}_{0} & = \mu_{B} \int \frac{dE\,dk_{y}}{4\pi^{2}} f(E_{f}) \mathrm{Im} (g^{\uparrow}_{i,i} - g^{\downarrow}_{i,i}) \boldsymbol{S} \\
& =\mu_{B} (\langle n_{i}^{\uparrow}\rangle - \langle n_{i}^{\downarrow}\rangle) \boldsymbol{S},
\end{aligned}
\end{equation}
\noindent where $\langle n_{i}^{\sigma}\rangle = \int \frac{dE}{\!2\pi} \, \rho_{i}^{\sigma}(E) f(E_{F})$ is the average number of $s$-electrons of spin $\sigma$ at atom $i$, and $\rho_{i}^{\sigma}=\mathrm{Im}\,\int\frac{dk_{y}}{\!2\pi}\,g_{i,i}^{\sigma}$ is the density of states (DOS). 

\par Let us define the charge and $z$-component of the spin currents given by Eqs.~(\ref{eq-I}) and (\ref{eq-Is}) in the absence of SOC with respect to the local coordinate frame aligned with the magnetization $\boldsymbol{S}$
\small
\begin{equation}
\begin{aligned}
\label{cur_ap}
I&=\frac{et}{\hbar}\int \frac{dE\,dk_{y}}{4\pi^{2}}[G^{<\uparrow\uparrow(0)}_{i+1,i}-G^{<\uparrow\uparrow(0)}_{i,i+1}\\
&\qquad\qquad\qquad+G^{<\downarrow\downarrow(0)}_{i+1,i}-G^{<\downarrow\downarrow(0)}_{i,i+1}],\\
I^{S_{z}}&=\int \frac{dE\,dk_{y}}{8\pi^{2}}[G^{<\uparrow\uparrow(0)}_{i+1,i}-G^{<\uparrow\uparrow(0)}_{i,i+1}\\
&\qquad\qquad\qquad-G^{<\downarrow\downarrow(0)}_{i+1,i}+G^{<\downarrow\downarrow(0)}_{i,i+1}].
\end{aligned}
\end{equation}
\normalsize
\noindent Using Eqs.~(\ref{Eq-G-col}) and (\ref{Eq-g-col}) we obtain
\small
\begin{equation}
\begin{aligned}
&I=\frac{et}{\hbar}\int \frac{dE\,dk_{y}}{4\pi^{2}}[\theta(4t^{2}-E^{\uparrow2})+\theta(4t^{2}-E^{\downarrow2})],\\
&I^{S_{z}}=\int \frac{dE\,dk_{y}}{8\pi^{2}}[\theta(4t^{2}-E^{\uparrow2})-\theta(4t^{2}-E^{\downarrow2})],
\end{aligned}
\end{equation}
\normalsize
\noindent where $\theta(t)$ is the Heaviside step function. Integration with respect to $k_{y}$ yields  $I=I^{\uparrow}+I^{\downarrow}$ and $I^{S_{z}}=\frac{\hbar}{2e}(I^{\uparrow}-I^{\downarrow})$, where
\begin{equation}
I^{\sigma}=\frac{e}{h}\int\limits_{-\infty}^{\infty}\frac{dE}{2\pi}(f_{L}-f_{R})D^{\sigma}(E) 
\end{equation}
\noindent and
\small
$$
\begin{aligned}
D^{\sigma}(E)&=\theta(16t^{2}-(E-\varepsilon^{\sigma})^{2})\left[\theta(4t^{2}-(E-\varepsilon^{\sigma}+2t)^{2})\right.\\
&+\frac{1}{\pi}\arccos{\left(\frac{E-\varepsilon^{\sigma}-2t}{2t}\right)}\theta\left(\frac{E-\varepsilon^{\sigma}}{t}\right)\\
&-\frac{1}{\pi}\arccos{\left(\frac{E-\varepsilon^{\sigma}+2t}{2t}\right)}\theta\left(-\frac{E-\varepsilon^{\sigma}}{t}\right)].
\end{aligned}
$$
\normalsize
\noindent Here $I^{\sigma}$ is the contribution of the charge current from the channel with spin $\sigma$, $D^{\sigma}$ is the corresponding transmission function, and $\varepsilon^{\sigma}=\varepsilon_{o}\pm j_{ex}$. Note that in the limit of low temperatures the integral for $D^{\sigma}$ can be taken analytically. 

\par Next, we collect the terms of Eq.~(\ref{eq-NEGF-final}) with the first power of $\lambda$
\small
\begin{equation}
\begin{aligned}
\hat{G}^{<(1)}_{i,i}&=f_{L}t\left(\hat{g}_{i,i}\hat{G}^{(1)+}_{i,i+1}+\hat{G}^{(1)}_{i,i}\hat{g}^{+}_{i,i+1}-\hat{g}_{i+1,i}\hat{G}^{(1)+}_{i,i}-\hat{G}^{(1)}_{i+1,i}\hat{g}^{+}_{i,i}\right)\\
&+f_{R}t\left(\hat{g}_{i,i}\hat{G}^{(1)+}_{i+1,i}+\hat{G}^{(1)}_{i+1,i}\hat{g}^{+}_{i,i}-\hat{g}_{i,i+1}\hat{G}^{(1)+}_{i,i}-\hat{G}^{(1)}_{i,i+1}\hat{g}^{+}_{i,i}\right)\\
&+if_{L}\lambda\left(\hat{g}_{i,i}\sigma_{y}\hat{g}_{i,i+1}^{+}+\hat{g}_{i+1,i}\sigma_{y}\hat{g}_{i,i}^{+}\right)\\
&-if_{R}\lambda\left(\hat{g}_{i,i}\sigma_{y}\hat{g}_{i+1,i}^{+}+\hat{g}_{i,i+1}\sigma_{y}\hat{g}_{i,i}^{+}\right),
\end{aligned}
\label{eq1111}
\end{equation}
\normalsize
\noindent where $\hat{G}^{(1)}$ stands for the GF's correction to first order in SOC
\begin{equation}
\hat{G}^{(1)}_{nm}(k_{y})=\frac{1}{2\pi}\int\limits_{-\pi}^{\pi} dk_{x}\, \hat{g}(\boldsymbol{k})\hat{\mathcal{H}}_{SO}(\boldsymbol{k})\hat{g}(\boldsymbol{k})e^{ik_{x}(n-m)}.
\end{equation}
\noindent Having substituted $\hat{G}^{(1)}$ in Eq. (\ref{eq1111}), we obtain the final expression for the on-site matrix elements of $\hat{G}^{<(1)}_{i,i}$
\small
\begin{equation}
\begin{aligned}
&G_{i,i}^{<\uparrow\uparrow(1)}=i\lambda(f_{R}-f_{L})\left[\frac{1}{t}\,\mathrm{Re}(g_{i,i}^{\uparrow}+g_{i,i}^{\downarrow})\sin{\theta}\cos{\phi}\right. \\
&+\left.\frac{1}{2}\mathrm{Re}[\Lambda_{1}+\Lambda_{2}-\frac{1}{t}(g_{i,i}^{\uparrow}+g_{i,i}^{\downarrow})]\sin{2\theta}\sin{\phi}\right],\\
&G_{i,i}^{<\uparrow\downarrow(1)}=\lambda(f_{R}-f_{L})[(\sin^{2}{\theta}\sin^{2}{\phi}-1)\mathrm{Re}[\Lambda_{1}+\Lambda_{2}]\\
&+\frac{i}{2}\mathrm{Re}[\Lambda_{1}+\Lambda_{2}-\frac{1}{t}(g_{i,i}^{\uparrow}-g_{i,i}^{\downarrow})]\sin^{2}{\theta}\sin{2\phi}\\
&-\frac{1}{t}\mathrm{Re}(g_{i,i}^{\uparrow}+g_{i,i}^{\downarrow})]\sin^{2}{\theta}\sin^{2}{\phi}],\\
&G_{i,i}^{<\downarrow\uparrow(1)}=\lambda(f_{R}-f_{L})[(1-\sin^{2}{\theta}\sin^{2}{\phi})\mathrm{Re}[\Lambda_{1}+\Lambda_{2}]\\
&+\frac{i}{2}\mathrm{Re}[\Lambda_{1}+\Lambda_{2}-\frac{1}{t}(g_{i,i}^{\uparrow}-g_{i,i}^{\downarrow})]\sin^{2}{\theta}\sin{2\phi}\\
&+\frac{1}{t}\mathrm{Re}(g_{i,i}^{\uparrow}+g_{i,i}^{\downarrow})]\sin^{2}{\theta}\sin^{2}{\phi}],\\
&G_{i,i}^{<\downarrow\downarrow(1)}=i\lambda(f_{R}-f_{L})\left[\frac{1}{t}\,\mathrm{Re}(g_{i,i}^{\uparrow}+g_{i,i}^{\downarrow})\sin{\theta}\cos{\phi}\right. \\
&-\left.\frac{1}{2}\mathrm{Re}[\Lambda_{1}+\Lambda_{2}-\frac{1}{t}(g_{i,i}^{\uparrow}+g_{i,i}^{\downarrow})]\sin{2\theta}\sin{\phi}\right],
\end{aligned}\label{eq-nefg-matrix}
\end{equation}
\normalsize
\noindent where
$$
\begin{gathered}
\Lambda_{1}=g_{i,i}^{\uparrow*}(g_{i+1,i}^{\downarrow}-2tK),\\
\Lambda_{2}=g_{i,i}^{\downarrow*}(g_{i+1,i}^{\uparrow}-2tK),\\
K=\frac{1}{2\pi}\int\limits_{-\pi}^{\pi}dk_{x}\,g^{\uparrow}(\boldsymbol{k})g^{\downarrow}(\boldsymbol{k})\sin^{2}{k_{x}}.
\end{gathered}
$$
\noindent Since all matrix elements of $\hat{G}^{<(1)}$ are proportional to the difference between the Fermi-Dirac functions of the left and right leads, they vanish in equilibrium. Following the definition of Eq.~(\ref{eq-M}), we can write the current-induced contribution $\boldsymbol{\mu}_{1}$ to the magnetic moment to first order in~$\lambda$
\begin{equation}
\begin{gathered}
\mu_{1x} = \mu_1 \sin^{2}{\theta}\sin{2\phi},\\
\mu_{1y} = 2 \mu_1 (\sin^{2}{\theta}\sin^{2}{\phi}-1),\\
\mu_{1z} = \mu_1 \sin{2\theta}\sin{\phi},
\end{gathered}
\end{equation}
\noindent where the coefficient $\mu_1$ depends on the band structure and applied bias
\begin{equation}
\mu_1 = - \frac{\mu_{B}\lambda}{4\pi^{2}} \int dE dk_{y} (f_{L}-f_{R}) \mathrm{Re}[ \Lambda_{1} + \Lambda_{2}].\label{eq-mu1}
\end{equation}
\noindent It can be further simplified
$$
\mathrm{Re}[\Lambda_{1}+\Lambda_{2}]=\frac{1}{2tj_{ex}}\left(\theta(4t^{2}-E^{\uparrow2})-\theta(4t^{2}-E^{\downarrow2})\right),
$$
\noindent that yields
\begin{equation}
\begin{aligned}
\mu_1 & = - \frac{\mu_{B}\lambda}{8\pi^{2} tj_{ex}} \int dE (f_{L}-f_{R}) ( D^{\uparrow}(E) - D^{\downarrow}(E) ) \\
& = - \frac{\mu_{B}\lambda\hbar}{2etj_{ex}} (I^{\uparrow}-I^{\downarrow}).
\end{aligned}
\end{equation}

\par Finally, the total magnetic moment $\boldsymbol{\mu}$ to first order in SOC can be written
\begin{equation}
\boldsymbol{\mu} = \boldsymbol{\mu}_{\parallel} + \boldsymbol{\mu}_{\perp} = \boldsymbol{S} (\mu_{0} +2 S_{y} \mu_1) + (0, -2 \mu_1, 0),
\end{equation}
\noindent where only the second term leads to SOT, which is of the field-like symmetry,
\begin{equation}
\boldsymbol{T} = T_{\perp} \, (S_{z}, 0, -S_{x})
\end{equation}
\noindent and
\begin{equation}
T_{\perp} = -\frac{2j_{ex}\mu_1}{\mu_{B}} = \frac{\hbar\lambda}{et} (I^{\uparrow} - I^{\downarrow}).
\end{equation}

\section{Equation of motion for the spin density}
\noindent Generally, the result given above can be derived directly from the equation of motion for the spin density. The spin current operator in the second quantization is
\begin{equation}
\hat{\boldsymbol{j}}^{S}_{i\rightarrow j}=-\frac{i}{4}\sum_{\sigma\sigma'}\left(\hat{c}_{j}^{+\sigma'}\{\boldsymbol{\sigma},\widetilde{t}_{ji}\}^{\sigma'\sigma}\hat{c}_{i}^{\sigma}-\mathrm{H.C.}\right),
\end{equation}
where $\{\boldsymbol{\sigma},\widetilde{t}_{ji}\}$ is the symmetrized product of the Pauli matrices and tight-binding Hamiltonian of a general form
$$
\hat{\mathcal{H}}=\sum_{ij,\sigma\sigma'}\widetilde{t}_{ij}^{\sigma\sigma'}\hat{c}_{i}^{+\sigma}\hat{c}_{j}^{\sigma'}.
$$
\noindent In the absence of SOC, the Hamiltonian $\hat{\mathcal{H}}_{0}$ given by Eq.~(\ref{Eq-Ho}) produces the well-known ``kinetic'' contribution to the spin current
\begin{equation}
\begin{aligned}
\hat{j}^{S_{x}}_{i\rightarrow j}&=-\frac{it}{2}(\hat{c}_{j}^{+\uparrow}\hat{c}_{i}^{\downarrow}+\hat{c}_{j}^{+\downarrow}\hat{c}_{i}^{\uparrow}-\hat{c}_{i}^{+\uparrow}\hat{c}_{j}^{\downarrow}-\hat{c}_{i}^{+\downarrow}\hat{c}_{j}^{\uparrow}),\\
\hat{j}^{S_{y}}_{i\rightarrow j}&=\,-\frac{t}{2}\,(\hat{c}_{j}^{+\uparrow}\hat{c}_{i}^{\downarrow}-\hat{c}_{j}^{+\downarrow}\hat{c}_{i}^{\uparrow}-\hat{c}_{i}^{+\uparrow}\hat{c}_{j}^{\downarrow}+\hat{c}_{i}^{+\downarrow}\hat{c}_{j}^{\uparrow}),\\
\hat{j}^{S_{z}}_{i\rightarrow j}&=-\frac{it}{2}(\hat{c}_{j}^{+\uparrow}\hat{c}_{i}^{\uparrow}-\hat{c}_{j}^{+\downarrow}\hat{c}_{i}^{\downarrow}-\hat{c}_{i}^{+\uparrow}\hat{c}_{j}^{\uparrow}+\hat{c}_{i}^{+\downarrow}\hat{c}_{j}^{\downarrow}),
\end{aligned}
\end{equation}
\noindent while the SOC part of the Hamiltonian $\hat{\mathcal{H}}_{SO}$ given by Eq.~(\ref{Eq-Hsoc}) gives rise to the SOC-induced spin currents with the only non-zero components
\begin{equation}
\begin{aligned}
\hat{j}^{S_{x}}_{SO,i\rightarrow i + \boldsymbol{e}_{x}} & = \lambda \hat{\rho}_{i, i + \boldsymbol{e}_{x}}, \\
\hat{j}^{S_{y}}_{SO,i\rightarrow i + \boldsymbol{e}_{y}} & = - \lambda \hat{\rho}_{i, i + \boldsymbol{e}_{y}}, 
\end{aligned}
\end{equation}
\noindent where $\hat{\rho}_{ij} = \frac{1}{2}\sum_{\sigma} (\hat{c}_{j}^{+\sigma}\hat{c}_{i}^{\sigma} + \mathrm{H.C.})$.

\par Starting from the equation of motion for the spin density operator in a Heisenberg picture
\begin{equation}
\frac{d\hat{\boldsymbol{s}}_{i}}{dt} = - \frac{i} {\hbar} [ \hat{\boldsymbol{s}}_{i},\ \hat{\mathcal{H}}]
\end{equation}
\noindent with $\hat{\boldsymbol{s}}_{i} = \frac{\hbar}{2} \sum_{\sigma\sigma'} \hat{c}_{i}^{+\sigma'} \boldsymbol{\sigma}^{\sigma'\sigma} \hat{c}_{i}^{\sigma}$, we obtain \cite{Nikolic}
\begin{equation}
\frac{d\hat{\boldsymbol{s}}_{i}}{dt}+\mathrm{div}\,\hat{\boldsymbol{j}}^{S}_{i}-j_{ex}\hat{\boldsymbol{s}}_{i}\times\boldsymbol{S}=\hat{\boldsymbol{j}}^{\omega}_{i},
\end{equation}
\noindent where $\hat{\boldsymbol{j}}^{\omega}$ is given by
\begin{equation}
\begin{aligned}
\hat{j}^{\omega_{x}}_{i} & = \frac{\lambda}{t} (\hat{j}^{S_{z}}_{i\rightarrow i+\boldsymbol{e}_{x}} + \hat{j}^{S_{z}}_{i-\boldsymbol{e}_{x}\rightarrow i}), \\
\hat{j}^{\omega_{y}}_{i} & = \frac{\lambda}{t} (\hat{j}^{S_{z}}_{i\rightarrow i + \boldsymbol{e}_{y}} + \hat{j}^{S_{z}}_{i - \boldsymbol{e}_{y}\rightarrow i}), \\
\hat{j}^{\omega_{z}}_{i} & = \frac{\lambda}{t} (\hat{j}^{S_{x}}_{i\rightarrow i + \boldsymbol{e}_{x}} + \hat{j}^{S_{x}}_{i-\boldsymbol{e}_{x} \rightarrow i} + \hat{j}^{S_{y}}_{i \rightarrow i + \boldsymbol{e}_{y}} + \hat{j}^{S_{y}}_{i - \boldsymbol{e}_{y} \rightarrow i})
\end{aligned} \label{Eq-Iw}
\end{equation}
\noindent and reflects the fact that spin is not a conserved quantity in the presence of SOC which acts as a magnetic field forcing spin to precession.\cite{Sun}

\par Taking statistical averages of the spin density in a steady sate gives
$$
\left\langle \mathrm{div} \, \hat{\boldsymbol{j}}^{S}_{i} \right\rangle - j_{ex} \left\langle \hat{\boldsymbol{s}}_{i} \times \boldsymbol{S} \right\rangle = \left\langle \hat{\boldsymbol{j}}^{\omega}_{i} \right\rangle.
$$
\noindent For the sake of simplicity, we consider the magnetization lying in the $xz$ plane, $\boldsymbol{S} = (\sin{\theta}, 0, \cos{\theta})$, and the spin polarized current flowing along the $x$ axis. In a ballistic regime, the divergence of the spin current is close to zero on a macroscopic scale, when the system is large enough so that all inhomogeneities are negligible.  To first order in SOC, we can also neglect all the induced currents flowing along the transverse $y$ direction. However, one should take into account that the non-collinearity between $\hat{\boldsymbol{s}_{i}}$ and $\boldsymbol{S}$ to first order in SOC is driven by the transverse component of the spin current, that is $\hat{j}^{S_{z}}_{i\rightarrow i + \boldsymbol{e}_{y}}$, and according to Eq.~(\ref{Eq-Iw}) only $\hat{s}_{y}$ produces SOT with the non-zero $x$ and $z$ components
\begin{equation}
\begin{aligned}
-j_{ex} \left\langle \hat{\boldsymbol{s}}_{y}\right\rangle S_{z} & = \frac{2\lambda}{t} \left\langle \hat{j}^{S_{z}}_{i\rightarrow i+\boldsymbol{e}_{x}} \right\rangle, \\
j_{ex}\left\langle \hat{\boldsymbol{s}}_{y}\right\rangle S_{x} & = \frac{2\lambda}{t} \left\langle \hat{j}^{S_{x}}_{i\rightarrow i+\boldsymbol{e}_{x}} \right\rangle,
\end{aligned}
\end{equation}
\noindent respectively. All the averages in this equation can be expressed through NEGF $\hat{G}^{<}_{i,j} = i \langle \hat{c}^{+}_{j}\hat{c}_{i}^{} \rangle$. Taking into account Eqs.~(\ref{eq-M}) and (\ref{eq-T}), and substituting $\langle \hat{j}^{S_{z}}_{i\rightarrow i+\boldsymbol{e}_{x}}\rangle=I^{S_{z}}\cos{\theta}$ and $\langle \hat{j}^{S_{x}}_{i\rightarrow i+\boldsymbol{e}_{x}}\rangle=I^{S_{z}}\sin{\theta}$ after a transformation to the global coordinate frame, we finally obtain
\begin{equation}
\begin{aligned}
\boldsymbol{T}_{\perp} & = \frac{2\lambda}{t} I^{S_{z}} (S_{z},0,-S_{x}) \\
& = \frac{\hbar\lambda}{et}(I^{\uparrow}-I^{\downarrow})(S_{z},0,-S_{x}),
\end{aligned}
\end{equation}
\noindent that is the same result obtained in Appendix A. It is worth noting that the DLT is absent in this derivation as the electric does not appear explicitly in the Hamiltonian.

\end{document}